\UseRawInputEncoding
\documentclass[aps, 
floats,
preprint, 
showpacs,
nofootinbib,
tightenlines,
floatfix
]{revtex4}%
\usepackage{eurosym}
\usepackage{amsfonts}
\usepackage{amsmath}
\usepackage{array}
\usepackage{graphicx}
\usepackage{subfigure}
\usepackage{relsize}
\usepackage{amssymb}%
{\catcode`\|=\active \gdef\Braket#1{\left<\mathcode`\|"8000\let|\bravert
{#1}\right>}}
\def\bravert{\egroup\,\vrule\,\bgroup}

\newcommand{\beq}{\begin{eqnarray}}
\newcommand{\eeq}{\end{eqnarray}}
\newcommand{\bw}{\begin{widetext}}
\newcommand{\ew}{\end{widetext}}
\newcommand{\open}{\sphericalangle}

\begin{document}

\title{Insights into the higher-twist distribution $e(x)$ at CLAS}
\author{A. Courtoy}
\email{Aurore.Courtoy@ulg.ac.be}
\affiliation{
IFPA, AGO Department, Universit\'e de Li\`ege, B‰t. B5, Sart Tilman B-4000 Li\`ege, Belgium
\\INFN, Laboratori Nazionali di Frascati,
Via E. Fermi, 40, I-00044 Frascati (Roma), Italy
}

\date{\today }

\begin{abstract}
This preprint has been superseded by Ref.~\cite{Courtoy:2022kca}. The latter should be consulted and referred to for the extraction of the $e(x)$ PDF.\\

We present the extraction of the twist-3 PDF, $e(x)$, through the  analysis of the  preliminary data for the $\sin\phi$-moment of the beam-spin asymmetry for  di-hadron Semi-Inclusive DIS at CLAS at 6 GeV. Pion-pair production off unpolarized target in the DIS regime provide an access to the higher-twist Parton Distribution Functions $e(x)$ and to  Di-hadron Fragmentation Functions. The latter have been extracted from the semi-inclusive production of two hadron pairs in back-to-back jets in $e^+e^-$ annihilation at Belle.
The  $e(x)$ PDF offers important insights into the physics of the largely-unexplored quark-gluon correlations, and its $x$-integral is related to the marginally-known
scalar-charge of the nucleon, and to the pion-nucleon $\sigma$-term, a fundamental property of the nucleon. 
\end{abstract}

\pacs{}
\maketitle

\section{ Introduction}
Hard processes are described in QCD by envisaging a perturbative stage (pQCD) where a hard collision involving quark and gluons occurs, followed by a non-perturbative stage characterizing hadron structure.
For example, in  Deep Inelastic Scattering (DIS)  
the hard scattering part of the process, $\gamma^* q \rightarrow X$, occurs at very short light cone distances, or for small configurations of quarks and gluons  which can be presently described within pQCD.
The large distance contribution is parameterized in terms of Parton Distribution Functions (PDF), which contain the structural information on the target. Formally, this factorization can be achieved in an Operator Product Expansion (OPE) style. In the collinear approach, the leading order of the non-perturbative contributions, called leading-twist (twist-2), is composed of three PDFs depending only on the fraction $x$ of the longitudinal momentum of the target and on the photon virtuality. The subleading-twist is equivalently composed of three PDFs.

The experimental determination of collinear structure of the proton is  not complete. At leading-twist, only the unpolarized PDF, $f_1(x)$, is well known. The helicity distribution, $g_1(x)$, is less constrained, while the transversity distribution $h_1(x)$ is known to some extent. The subleading--twist picture consists in three collinear PDFs, $e(x), h_L(x)$ and $g_T(x)$. While  these functions provide direct and unique insights into the dynamics inside
hadrons~\cite{Jaffe:1989xx}, experimental information is still scarce~\cite{Airapetian:2011wu}. In particular, the chiral-odd twist-3 PDF $e(x)$ encloses important knowledge on the largely unexplored quark-gluon 
correlations.  In general, higher-twist PDFs describe multiparton distributions 
corresponding to the interference of higher Fock components in the hadron wave functions, 
and as such have no probabilistic partonic interpretations. Yet they offer fascinating 
doorways to studying the structure of the nucleon.
Higher-twist  contributions are also indispensable to correctly extract twist-2 components from data.
Although suppressed with respect to twist-2 observables by $1/Q$, twist-3 observables are not small in the kinematics of 
fixed target experiments. The CLAS experiment, installed in the Hall-B of the Jefferson Laboratory, represents the ideal environement to study ${\cal O}(1/Q)$ contributions, thanks to the low average photon virtuality $Q^2$ explored
in its experiments, and its capability to extract the observables of interest in a wide kinematic range.

The golden channel to access $e(x)$ is Semi-Inclusive production of pion pairs in the Deep Inelastic regime.  
In di-hadron SIDIS Single-Spin Asymmetries the PDF $e(x)$ appears coupled to the {\it chiral-odd} Interference Fragmentation Function $H_1^{\open}$~\cite{Bacchetta:2003vn}, that, together with the unpolarized Di-hadron Fragmentation Function (DiFF) $D_1$, constitutes a crucial ingredient to obtain information on 
PDFs. Such a
process can be analyzed in the framework of the collinear factorization, making the di-hadron SIDIS a unique tool to study the higher-twist effects appearing as $\sin\phi$
modulations in beam-spin dependent azimuthal moments of the SIDIS cross section.
The  Interference Fragmentation Function $H_1^{\open}$ has been recently extracted \cite{Courtoy:2012ry} from Belle data \cite{Vossen:2011fk}, providing an important ingredient toward the PDF extraction.
As to $D_1$, in the absence of data for di-hadron multiplicities related to the unpolarized DiFF, it has been fitted to the output of PYTHIA~\cite{Sjostrand:2003wg} tuned for Belle kinematics. The DiFF framework has proven its efficiency in the transverse target case, leading to the first extraction  of the collinear transversity PDF, $h_1(x)$,  for HERMES and COMPASS data~\cite{Bacchetta:2011ip,Bacchetta:2012ty}. 

In this paper we present an extraction of the higher-twist PDF $e(x)$.
An extended study and review on the chiral-odd parton distribution has been published 10 years ago~\cite{Efremov:2002qh}. A first attempt to access the $e(x)$ PDF was proposed in Ref.~\cite{Efremov:2002ut} through the analysis of the single-hadron SIDIS
Beam-Spin Asymmetry measured by the CLAS Collaboration \cite{Avakian:2003pk}, which involves TMD factorization and four terms in the structure function. Recent data for the Beam Spin Asymmetry for single-pion semi-inclusive electro-production~\cite{Airapetian:2006rx,Gohn:2014zbz} should bring more light on the TMD $e(x, k_T)$.\\

The paper is organized as follows.
In Section~\ref{sec:HT}, we describe the higher-twist physics, especially $e(x)$. Section~\ref{sec:BSA} is devoted to the framework for di-hadron Beam Spin Asymmetry (BSA).  In Section~\ref{sec:result}, we present the analysis and extraction of $e(x)$. The results are discussed. We then conclude. 

\section{partonic quantities}
\label{sec:HT}

\subsection{The chiral-odd twist-3 $e(x)$}

The chiral-odd $e(x)$ twist-3 distribution is defined as
	\begin{eqnarray}
	e^q(x)&=&\frac{1}{2M}\, \int \frac{d\lambda}{2\pi}\, e^{i\lambda x}\langle P| \bar{\psi}_q(0)\psi_q(\lambda n)|P\rangle\quad,
	\label{eq:def}
	\end{eqnarray}
	for quarks, and for antiquarks, $e^{\bar{q}}(x)=e^q(-x)$. $n$ is a light-like vector ; Eq.~(\ref{eq:def}) is expressed in the light-cone gauge, {\it i.e.} where the gauge link becomes unity.
	
Twist-3 PDFs are suppressed in the OPE expansion by a factor $M/P^+$ {\it w.r.t.} the twist-2~PDFs.
The origin of that suppression can be either kinematical, dynamical or due to quark mass terms. The separation between these three contributions comes from a QCD operator identity for the
  non-local quark-quark operator, $\bar\psi\psi$,~\cite{Jaffe:1991kp,Jaffe:1991ra,Balitsky:1996uh,Belitsky:1997zw,Koike:1996bs}. 
%
%
%
 Kinematical twist-3 can be reduced to an expression containing only  twist-2 PDFs via QCD equations of motion, it is the so-called Wandzura-Wilczek  (WW) approximation~\cite{Wandzura:1977qf}.
  The PDF $e(x)$ vanishes in this  approximation. QCD equations of motion allow to decompose the chiral-odd twist-3 distributions into 3 terms,
\begin{eqnarray}
\nonumber\\
e^q(x)&=&e_{\mbox{\tiny{loc}}}^q(x)+e_{\mbox{\tiny{gen}}}^q(x)+e_{\mbox{\tiny{mass}}}^q(x)\quad. \label{eq:def_ex}\\
\nonumber
\end{eqnarray}
The first term comes from the {\it local} operator: 
\begin{eqnarray}
e_{\mbox{\tiny{loc}}}^q(x)&=&\frac{1}{2M}\, \int \frac{d\lambda}{2\pi}\, e^{i\lambda x}\langle P| \bar{\psi}_q(0)\psi_q(0)|P\rangle=\frac{\delta(x)}{2M}\langle P| \bar{\psi}_q(0)\psi_q(0)|P\rangle\quad;
\end{eqnarray}
the second term is a dynamical or {\it genuine} twist-3 contribution, {\it e.g.} it is interaction dependent and contains explicit gluon fields; the last term is proportional to the quark mass and  its  Mellin moments are expressed as
\begin{eqnarray}
\int_{-1}^1 dx \, x^{n-1}  e_{\mbox{\tiny{mass}}}^q(x)&=&\frac{m_q}{M}\, \int_{-1}^1 dx \, x^{n-2}  f_1^q(x)\quad,
\end{eqnarray}
for $n>1$ and is zero for $n=0$.

The QCD evolution of $e(x)$ has been studied up to NLO~\cite{Balitsky:1996uh,Belitsky:1997zw,Koike:1996bs}. Due to the chiral-odd nature of the current, there is no mixing with gluons. Evolution of twist-3 operators is complex but can be reduced to a DGLAP-like scheme in the large-$N_c$ limit.
  
The PDF $e(x)$ has been calculated in various models. We cite the chiral quark soliton model, {\it e.g.}~\cite{Ohnishi:2003mf,Schweitzer:2003uy},  the MIT bag model~\cite{Jaffe:1991ra,Avakian:2010br},  the spectator model~\cite{Jakob:1997wg,simo_abha},  the instanton QCD vacuum calculus  and the perturbative light-cone Hamiltonian approach to ${\cal O}(\alpha_s)$ with a quark target~\cite{Burkardt:2001iy,Mukherjee:2009uy}. 
In Ref.~\cite{Schweitzer:2003uy} the non-relativistic limit of $e^q(x)$ was studied. A calculation in the light-front quark model is ongoing~\cite{barbara_ongoing}.
\\


The chiral-odd twist-3 PDF $e(x)$ carries important hadronic information. It offers a unique road to the determination of the scalar charge, {\it i.e.}
the first Mellin moment of $e(x)$:
\begin{eqnarray}
\nonumber\\
\int_{-1}^1 dx\, e^q(x, Q^2)&=&\int_{-1}^1 dx\,e_{\mbox{\tiny{loc}}}^q(x, Q^2)=\frac{1}{2M}\, \langle P| \bar{\psi}_q(0)\psi_q(0)|P\rangle (Q^2)=\sigma_q(Q^2)\quad.\label{eq:scalarcharge}\\
\nonumber
\end{eqnarray}
 The isoscalar combination  of the scalar charge is related to the pion-nucleon $\sigma$-term%
\begin{eqnarray}
\sigma_u (Q^2)\,+\,\sigma_d (Q^2)
&\equiv& \frac{\sigma_{\pi N}}{\left(m_u(Q^2)+m_d(Q^2)\right)/2}\quad.
\end{eqnarray}
The pion-nucleon $\sigma$-term is normalization point invariant. 
It is  related to the strangeness content of the proton. 
%
%
%
%
The $\sigma$-term represents the contribution from the finite quark masses to the mass of the nucleon~\cite{Ji:1994av}. The  value $\sigma_{\pi N}=79\pm7$ MeV  was obtained in Ref.~\cite{Pavan:2001wz}.

Besides being  fundamental characteristics of the nucleon, the scalar charges might be important in the search for physics Beyond the Standard Model.
For instance, in a study of the elastic scattering of supersymmetric cold dark matter particles on nucleons, it has been shown that the cross sections depend strongly on the value of the pion-nucleon $\sigma$-term~\cite{Ellis:2008hf}. General model-independent bounds on direct dark matter detection include all possible effective operators, beyond the $V-A$ electroweak structure~\cite{DelNobile:2013sia}. A classification of these operators and their implications include scalar form factors, that are related to the scalar charges in the forward limit.
Also, the isovector scalar charge is related to ``new currents" in beta decays, in the sense that the leptonic current allows the weak $V-A$ current structure in the Standard Model. New structures, such as scalar and tensor, would give hint of physics Beyond the Standard Model~\cite{Bhattacharya:2011qm} if detected.

The sum rule in Eq.~(\ref{eq:scalarcharge}) is not strickly speaking related to  a charge, as that charge is not scale invariant. %
Moreover the contribution to the charges comes only from the singular --{\it local}-- part of the twist-3 PDF. While little can be told experimentally on the singular contribution, it has been studied in various models.
 In chiral models, the presence of this singular term in the distribution is inseparably connected with the nonzero value of quark condensate in the spontaneously-breaking QCD vacuum~\cite{Schweitzer:2003uy,Efremov:2002qh,Wakamatsu:2003uu}. 

The second moment of $e^q$ is proportional to the number of valence quarks of flavor $q$,
\begin{eqnarray}
\int_{-1}^1 dx\, x e^q(x)&=&\int_{0}^1 dx\, x (e^q- e^{\bar{q}})(x)=\frac{m_q(Q^2)}{M}\, N_q\quad,
\label{eq:sndmmt_e}
\end{eqnarray}
and it vanishes in the chiral limit.

The third moment of the chirally odd twist-3  parton distribution involves the {\it genuine} part and  can be related to the transverse force experienced by a transversely polarized quark ejected from a transversely polarized nucleon~\cite{Burkardt:2008ps}.

  
  \subsection{Dihadron Fragmentation Functions}

The twist-2 $\pi^+\pi^-$-DiFFs are the unpolarized $D_{1}$ and the chiral-odd $H_{1}^{\open}$. The latter is $T$-odd. The $D_1^q$ is the 
unpolarized DiFF describing the hadronization of a parton with flavor $q$ into an unpolarized 
hadron pair plus anything else, averaging over the parton polarization. The $H_1^{\open\, q}$ 
is a chiral-odd DiFF describing the correlation between the transverse polarization of the 
fragmenting parton with flavor $q$ and the azimuthal orientation of the plane containing the 
momenta of the detected hadron pair. 
In a Partial Wave Analysis (PWA), the physical interpretation of the dominant contribution to $H_1^{\open}$ is related to the interference between relative $p$ and $s$ wave of the pion-pairs, while, for $D_1$, the pion-pairs  are in  relative $s$ waves~\cite{Bacchetta:2002ux}. 

DiFFs depend on the fraction of longitudinal momentum, $z=z_1+z_2$, of fragmenting quark carried by the pion-pair,  on the ratio $\zeta=(z_1-z_2)/z$ ---that can be expressed in terms of the polar angle $\theta$, formed bewteen the direction of the back-to-back emission of the two hadrons  in the center of mass frame and the direction of average momentum of the hadron pair in the target rest frame--- and on the invariant mass of the pair, $m_{\pi\pi}$~\cite{Bacchetta:2002ux}.

DiFFs have been studied in models~\cite{Bacchetta:2006un,Casey:2012ux,Matevosyan:2014gea} and have been analyzed  for $\pi^+\pi^-$ production from Belle data~\cite{Courtoy:2012ry}. In particular, $H_1^{\open}$ was extracted from the Artru-Collins asymmetry measured at Belle, using $D_1$ fitted from the output of the MonteCarlo event generator tuned for Belle~\cite{Courtoy:2012ry}.
A functional form at the hadronic scale $Q_0^2=1$ GeV$^2$ was found, fitting the $100$ GeV$^2$ data. The  range of validity of the DiFF fits reflects the kinematic range of the Belle data. In particular, the  integrated range in invariant mass considered for the fit is limited to $2 m_\pi \leq m_{\pi\pi} \leq 1.29$ GeV,  the upper cut excluding scarcely populated or frequently empty bins for the Artru-Collins asymmetry. This limit varies bin by bin and the upper limit in $m_{\pi\pi}$ can be as low as $0.9$ GeV for $z=0.25$.\footnote{See Fig.~6 of Ref.~\cite{Courtoy:2012ry} and Ref.~\cite{Vossen:2011fk}.}
\\

At twist-3, the number of DiFFs increase. In particular there are four {\it genuine}  twist-3 DiFFs, $\widetilde{D}^{\open},\, \widetilde{G}^{\open}, \, \widetilde{E}$ and $\widetilde{H}$~\cite{Bacchetta:2003vn}.  The functions $\widetilde{D}^{\open},\, \widetilde{G}^{\open}$ are also Interference Fragmentation Functions, like $H_{1}^{\open}$. The genuine twist-3 DiFFs  describe the fragmentation of a quark, the propagator of which is corrected by gluon fields up to order ${\cal O}(1/Q)$.  They vanish in the Wandzura-Wilzcek approximation.
Up to date, there is no clear experimental information about higher-twist DiFFs.


\section{Beam-spin asymmetry in SIDIS off proton target}
\label{sec:BSA}

We consider the process
\begin{equation}
  \label{2hsidis}
\ell(l) + N(P) \to \ell(l') + h_1(P_1)+h_2(P_2) + X ,
\end{equation}
where $\ell$ denotes the beam lepton, $N$ the nucleon target,  $h_1$ and $h_2$ the produced hadrons, 
and where four-momenta are given in parentheses. 
We work in the one-photon exchange approximation and neglect the lepton mass.   
The momentum transferred to the nucleon target is $q=l-l'$.
The masses of the of final hadrons are 
$m_1$, $m_2$ and  their momenta  are, respectively, $P_1$, 
$P_2$. The total momentum of the pair is $P_h=P_1+P_2$ ; the relative momentum $R=(P_1-P_2)/2$ and its component   orthogonal to $P_h$ is  $R_T \equiv R -(R\cdot\hat{P}_h)\hat{P}_h$. 
The  
invariant mass squared of the hadron pair is $P_h^2 = m_{hh}^2$.  The SIDIS process is defined by the kinematic variables:
\begin{eqnarray}
&&x= \frac{Q^2}{2\,P\cdot q}\equiv x_B\quad ,\qquad y = \frac{P \cdot q}{P \cdot l}\quad, \qquad z = \frac{P \cdot P_h}{P\cdot q}=z_1+z_2\quad.
\label{eqn:kinem}
\end{eqnarray}
 The kinematics and the definition of the angles can be be found in, {\it e.g.}, Refs.~\cite{Bacchetta:2003vn,Bacchetta:2012ty}. We mention the azimuthal angle $\phi_R$ formed between the leptonic plane and the hadronic plane  identified by the vector $R_T$ and the virtual photon direction. The cross section for two particle SIDIS can be written in terms of modulations in the azimuthal angle $\phi_R$~\cite{Bacchetta:2006tn}.

In the limit $m_{hh}^2 \ll Q^2$ the structure functions of interest can be written in terms of PDFs and DiFFs, to leading-order, in the 
following way~\cite{Bacchetta:2003vn}:
\begin{align} 
\nonumber\\
\label{F_UUT}
\hspace{-3mm}
F_{UU ,T} & = \sum_q e_q^2\;x f_1^q(x)\, D_1^q\bigl(z,\cos \theta, m_{hh}\bigr), \phantom{\biggl[ \biggr]}
\\
\label{F_UUcosphi}
\hspace{-3mm}
F_{UU}^{\cos\phi_R} & =- \sum_q e_q^2\;x \frac{|\bm R| \sin \theta}{Q}\, \frac{1}{z}\, f_1^q(x)\,
\widetilde{D}^{\open\, q}\bigl( z,\cos \theta, m_{hh} \bigr), 
\\
\label{F_LUsinphi} 
\hspace{-3mm}
F_{LU}^{\sin\phi_R} &  = -\sum_q e_q^2\;x\frac{|\bm R| \sin \theta}{Q}\,
\biggl[
    \frac{M}{m_{hh}}\,x\, e^q(x)\, H_1^{\open\, q}\bigl(z,\cos \theta, m_{hh}\bigr)
    +\frac{1}{z}\,f_1^q(x)\,\widetilde{G}^{\open\, q}\bigl(z,\cos \theta, m_{hh}\bigr)\biggr],
\\
\hspace{-3mm}
\label{F_LL}
F_{LL}  & = \sum_q e_q^2\;x g_1^q(x)\, D_1^q\bigl(z,\cos \theta, m_{hh}\bigr), \phantom{\biggl[ \biggr]}
\\
\hspace{-3mm}
\label{F_LLcosphi}
F_{LL}^{\cos \phi_R} & =- \sum_q e_q^2\;x\frac{|\bm R| \sin \theta}{Q}\,
   \frac{1}{z}\,g_1^q(x)\,\widetilde{D}^{\open\, q}\bigl(z,\cos \theta, m_{hh}\bigr) , 
   \phantom{\biggl[ \biggr]}\\
   \nonumber
   \end{align} 
with the first subindex of the structure function corresponding to the beam polarization, the second to the target. 
We now consider the structure function $F_{LU}^{\sin\phi}$  in Eq.~(\ref{F_LUsinphi}) for $\pi^+\pi^-$ pair production. The relevant spin asymmetry can be built as ratios of structure functions. 
For the longitudinal polarization of the beam, {\it i.e.} the $LU$ combinations, one can define the following BSA: 
\begin{align} 
A_{LU}^{\sin \phi_R } \left( z, m_{\pi\pi}, x ; Q, y \right)
&=
\frac{\frac{4}{\pi}\sqrt{2\,\varepsilon (1-\varepsilon)} \int d \cos \theta \, F_{LU}^{\sin\phi_R}}
{\int d \cos \theta\, \left( F_{UU,T}+ \epsilon F_{UU,L} \right) }\quad,
\label{e:ssamaster}
\end{align} 
where $\varepsilon$ is the ratio of longitudinal and transverse photon flux and can be expressed in terms of $y$. 
%
%
Combining Eqs.~(\ref{F_UUT},\ref{F_LUsinphi}), to leading-order in $\alpha_s$ and leading term in the PWA, the BSA  becomes %
\begin{eqnarray}
&&A_{LU}^{\sin \phi_R } \left( x,  z, m_{\pi\pi} ; Q, y \right) \nonumber\\
&=&-\frac{W(y)}{A(y)}\,\frac{M}{Q}\,\frac{|\bf{R}  |}{m_{\pi\pi}} \, 
\frac{ \sum_q\, e_q^2\, \left[ x e^q(x, Q^2)\, H_{1, sp}^{\sphericalangle, q}(z, m_{\pi\pi}, Q^2)  + \frac{m_{\pi\pi}}{z M} \,  f_1^q(x, Q^2)\, 
          \tilde{G}_{sp}^{\sphericalangle, q}(z, m_{\pi\pi}, Q^2) \right]  } 
       { \sum_q\, e_q^2\,f_1^q(x, Q^2)\, D_{1,ss+pp}^q (z, m_{\pi\pi}, Q^2) }\quad ,\nonumber\\
 \label{eq:alu}
\end{eqnarray}
The dependence in $(z, m_{\pi\pi})$ is factorized in the DiFFs and kinematical factors, leaving the dependence in $x$ for the PDFs.
The twist-2 functions are $f_1(x), H_1^{\open} (z, m_{\pi\pi})$ and $D_1(z, m_{\pi\pi})$, while the twist-3 functions are $e(x)$ and $\tilde{G}^{\sphericalangle}(z, m_{\pi\pi})$.\footnote{From now on, we will drop the indices refering to the partial waves.}

\section{Result: extraction of the twist-3 PDF $e(x)$ }
\label{sec:result}

%
%
The longitudinal Beam-Spin Asymmetry $A_{LU}^{\sin \phi_R } \left( z, m_{\pi\pi}, x ; Q, y \right)$ in Eq.~(\ref{e:ssamaster}) has been recently extracted by the CLAS Collaboration on data collected by impinging the CEBAF $5.5$-GeV longitudinally-polarized
electron beam on an unpolarized $^2H$ hydrogen target \cite{BSA_silvia}. 

In Fig.~\ref{fig:triptic_zmh} the  measured asymmetry is shown in  two sets of 1D bins~\cite{BSA_silvia}, representing  respectively the $z$  and $m_{\pi\pi}$ dependence of the BSA. In Fig.~\ref{fig:e_extract_us_ww}  the $x$-dependence of the BSA  shows the data points used in the present extraction. The two plots of  Fig.~\ref{fig:triptic_zmh} are used to check the validity of the framework and its assumptions. \\
%
%
%
%
%
%

The  twist-3 chiral-odd PDF $e(x)$  is accessed through the $x$-dependent 1D projection of the BSA. The variables $(z, m_{\pi\pi})$, proper to the DiFFs,  do not enter in convolutions, so that, following the notation of Ref.~\cite{Bacchetta:2012ty}, we can define the following quantities:
\begin{eqnarray}
\nonumber\\
n_{q,\, i}(Q_i^2) &=& \int_{z_{\text{\tiny min},\,i}}^{z_{\text{\tiny max},\,i}} dz \,\int_{(m_{\pi\pi, \, \text{\tiny min}})_{i}}^{(m_{\pi\pi, \, \text{\tiny max}})_{i}}
 dm_{\pi\pi} \, D_1^q (z, m_{\pi\pi}; Q_i^2)  \; , \label{e:nq}  \\
n_{q,\, i}^\uparrow (Q_i^2) &=& \int_{z_{\text{\tiny min},\,i}}^{z_{\text{\tiny max},\,i}} dz \, \int_{(m_{\pi\pi, \, \text{\tiny min}})_{i}}^{(m_{\pi\pi, \, \text{\tiny max}})_{i}} dm_{\pi\pi} \, \frac{|\bm{R}|}{m_{\pi\pi}}\, H_{1}^{\open\, q}(z,m_{\pi\pi}; Q_i^2) \; , \label{e:nqup}\\
n_q^{\tilde{G}^{\sphericalangle}}  (Q_i^2) &=& \int_{z_{\text{\tiny min},\,i}}^{z_{\text{\tiny max},\,i}} dz \, \int_{(m_{\pi\pi, \, \text{\tiny min}})_{i}}^{(m_{\pi\pi, \, \text{\tiny max}})_{i}} dm_{\pi\pi} \, \frac{|\bm{R}|}{M}\, \tilde{G}^{\open\, q}(z,m_{\pi\pi}; Q_i^2) \; ,\\
\nonumber
\end{eqnarray}
where the index $i$ refers to the bin number and its respective integration limits. These integrated DiFFs  need to be  evaluated in the kinematical range of the experiment, which values are given in Tab.~\ref{tab:res}. 
The 1D projection of the BSA in Eq.~(\ref{eq:alu}) can be rewritten as:
\begin{eqnarray}
&&A_{LU}^{\sin \phi_R } \left( x_i,  m_{\pi\pi\,i,},  z_i ; Q_i, y_i \right) \nonumber\\
&=&-\frac{W(y_i)}{A(y_i)}\,\frac{M}{Q_i}\,
\frac{ \sum_q\, e_q^2\, \left[ x_i  e^q(x_i , Q_i ^2)\, n_{q,\, i}^{\uparrow}(Q_i^2)  + \frac{1}{ z_i  } \,  f_1^q(x_i , Q_i ^2)\, 
         n_{q,\, i}^{\tilde{G}^{\sphericalangle}} (Q_i^2) \right]  } 
       { \sum_q\, e_q^2\,f_1^q(x_i , Q_i ^2)\, n_{q, \, i} (Q_i^2) }\quad.
       \label{eq:alu_x}
\end{eqnarray}
The variables with subindex $i$ refer to average values for the bin $i$.\footnote{The integrated average differs from the bin average value only for bumpy distributions, {\it e.g.} the $m_{\pi\pi}$ behavior. We have checked that this difference was negligible for the unpolarized PDFs.} 

The twist-2 DiFFs are evaluated using the Pavia fit~\cite{Courtoy:2012ry} and the unpolarized PDFs, the MSTW08LO set~\cite{Martin:2009iq}. The $Q^2$ evolution of the twist-2 DiFFs 
has been properly included by using 
standard evolution equations in a collinear framework~\cite{Ceccopieri:2007ip} and by 
implementing leading-order (LO) chiral-odd splitting functions in the HOPPET code~\cite{Salam:2008qg}. The evolution is performed from $Q_0^2=1$ GeV$^2$ to the average scale of the bin $Q_i^2$. QCD evolution of the twist-3 DiFFs has not been studied yet.\footnote{Studies of the evolution of twist-3 fragmentation functions in  the multicolor limit show that there is no reciprocity in the anomalous dimensions for $e(x)$ and the twist-3 FF ${\cal G}_{\perp}^{\mbox{\tiny tw-3}}$~\cite{Belitsky:1997by}.} In this paper, evolution effects of the twist-3 DiFFs are assumed to be at most of  the order of magnitude of the running of $n_u^{\uparrow}(Q^2)$.

We use some approximations to further develop the sum over flavors ---we allow ourselves to $1-2 \%$ relative error on the DiFF contributions, which is negligible {\it w.r.t.} the  experimental error bars. The approximations are:
\begin{itemize}
\item The charm contribution to $f_1^{q=c}(x)$ is negligible {\it w.r.t} $q=u, \,d, \,s$ at JLab scales.
	
\item Invoking charge conjugation  yields to
\begin{eqnarray}
	D_{1}^{u\to\pi^+ \pi^-} &= &D_{1}^{\bar{u}\to\pi^+ \pi^-}\quad,\nonumber\\
	D_{1}^{d\to\pi^+ \pi^-}&=&D_{1}^{\bar{d}\to\pi^+ \pi^-}\quad ;
	\end{eqnarray}
together with isospin symmetry:	
	\begin{eqnarray}
	H_{1}^{\open u\to\pi^+ \pi^-} = - H_{1}^{\open d\to\pi^+ \pi^-} = - H_{1}^{\open \bar{u}\to\pi^+ \pi^-} = H_{1}^{\open \bar{d}\to\pi^+ \pi^-} \quad .
	\end{eqnarray}
	
\item The Interference FF for strange and charm is zero as there is no interference from sea quarks~\cite{Bacchetta:2006un}. 
	For $\tilde{G}^{\open}$ we expect the same relations as for $H_1^{\open}$.
\end{itemize}

The BSA in Eq.~(\ref{eq:alu_x}) becomes
\begin{eqnarray}
&&A_{LU}^{\sin \phi_R } \left( x_i,  m_{\pi\pi\,i,},  z_i ; Q_i, y_i \right)=-\frac{W(y_i)}{A(y_i)}\,\frac{M}{Q_i} \nonumber\\
&&\times\frac{  x_i  \left[ \frac{4}{9} e^{u_V}(x_i, Q_i^2)- \frac{1}{9} e^{d_V}(x_i, Q_i^2)\right] \,n_{u, i}^{\uparrow}(Q_i^2)  +   \left[\frac{4}{9} f_1^{u_V}(x_i , Q_i ^2)-\frac{1}{9} f_1^{d_V}(x_i , Q_i ^2)\right]/z_i\,
         n_{u, i}^{\tilde{G}^{\sphericalangle}} (Q_i^2)  } 
       {\sum_{q=u,d,s}\, e_q^2\,f_1^q(x_i , Q_i ^2)\, n_{q, \, i} (Q_i^2) }\quad.\nonumber\\
       \nonumber\\
       \nonumber\\
      && =-\frac{W(y_i)}{A(y_i)}\,\frac{M}{Q_i}\frac{x_i \left[e^V(x_i, Q_i^2)\right] n_{u,\, i}^{\uparrow}(Q_i^2)  + \left[f_1^V (x_i, Q_i^2) \right]/z_i\;
         n_{u,\, i}^{\tilde{G}^{\sphericalangle}} (Q_i^2) }{ \sum_{q=u,d,s}\, e_q^2\,f_1^q(x_i , Q_i ^2)\, n_{q, \, i} (Q_i^2)  }\quad,
       \label{eq:alu_x_fl}
\end{eqnarray}
where $f^{q_V}\equiv f^{q}-f^{\bar q}$. In Eq.~(\ref{eq:alu_x_fl}), we have defined the combinations $e^V$ and $f_1^{V}$.
The remaining unknown are then the twist-3 functions, $e(x)$ and $\tilde{G}^{\open}(z, m_{\pi\pi})$.
 While the twist-2 DiFFs are known, there is so far no study of the twist-3 DiFFs. A further assumption needs to be taken. In order to gain some insights onto the behavior of the genuine twist-3 DiFFs, we will define two extreme scenarios.
%

\subsection{ The Wandzura-Wilzcek scenario}

 In the Wandzura-Wilzcek approximation, the genuine twist-3 DiFFs vanish. This approximation is inspired by  the preliminary data on Double Spin Asymmetry (DSA) from CLAS, incorporating  the structure functions Eqs.~(\ref{F_UUcosphi},~\ref{F_LLcosphi}). These structure functions are expressed in terms of twist-2 PDFs and the twist-3 DiFF $\widetilde{D}^{\open}$, {\it i.e.} a genuine twist-3 DiFF that we expect to be bigger than $\widetilde{G}^{\open}$. These results indicate that the $\cos\phi$ modulation of the Double Spin Asymmetry (DSA)  is very small {\it w.r.t.} the constant term Eq.~(\ref{F_LL})~\cite{DSA_sergio}. 

\begin{figure}[t]
\begin{center}
		\includegraphics[scale= .6]{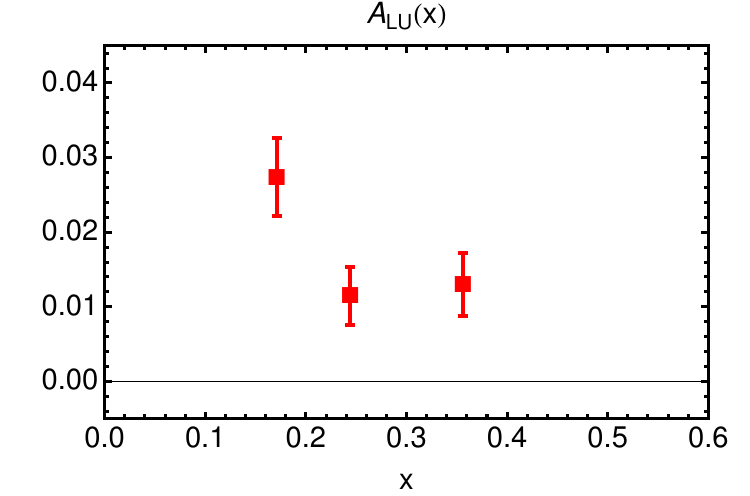}
		\includegraphics[scale= .6]{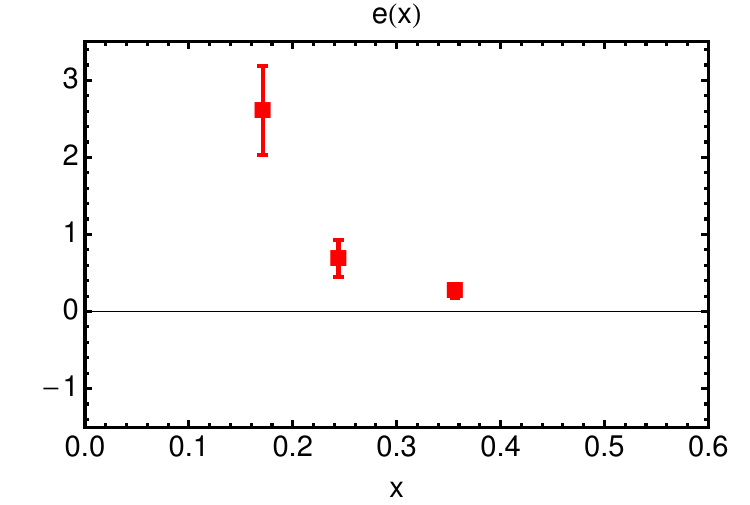}
\end{center}
\caption{  On the left panel, the $x$-dependent projection  of the preliminary BSA used to extract $e(x)$. On the  right panel, the extraction of the combination $e^V\equiv4 e^{u_V}(x_i, Q_i^2)/9-e^{d_V}(x_i, Q_i^2)/9$ in the WW scenario. The error bars correspond to the propagation of the experimental and DiFF errors. }
\label{fig:e_extract_us_ww}
\end{figure}

In this scenario, the 
 BSA Eqs.~(\ref{eq:alu_x_fl})  is straightfowardly inverted to get
\begin{eqnarray}
&&x_i^2\, e^{V}_{\mbox{\tiny WW}}(x_i, Q_i^2)= -\frac{A(y_i)}{W(y_i)}\frac{Q_i}{M}\, A_{LU}^{\sin \phi_R} \left( x_i,  m_{\pi\pi, \,i}, z_i ; Q_i^2, y_i\right)\nonumber\\
&&\times \frac{1}{9} \,\frac{ 4 x_i f_1^{u+\bar u} (x_i, Q_i^2)\,n_{u, i}(Q_i^2)+  x_i f_1^{d+\bar d}(x_i, Q_i^2)\, n_{d, i}(Q_i^2)+  x_i f_1^{s}(x_i, Q_i^2)\, n_{s, i}(Q_i^2)}{n_{u, i}^{\uparrow}(Q_i^2)}\quad.
\label{eq:therealextract}
\end{eqnarray}
The results are given in Tab.~\ref{tab:res} and shown in Fig.~\ref{fig:e_extract_us_ww}.
Notice that the range of integration in $m_{\pi\pi}$  goes beyond the range of known validity of the DiFF data set, {\it i.e.} the Belle data with $ 2m_{\pi}<m_{\pi\pi}<1.29$ GeV.
The error $\Delta\left(e^{V}(x)\right)$ reflects the propagation of the experimental --statistical and systematical-- error from Ref.~\cite{BSA_silvia} and the error on $H_1^{\open}$ taken from Ref.~\cite{Courtoy:2012ry}.
\\

%

%

\begin{figure}
\begin{center}
		\includegraphics[scale= .6]{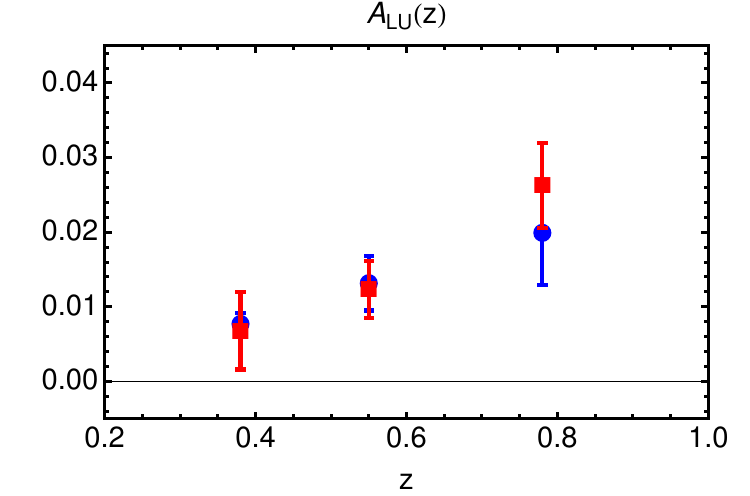}
		\includegraphics[scale= .6]{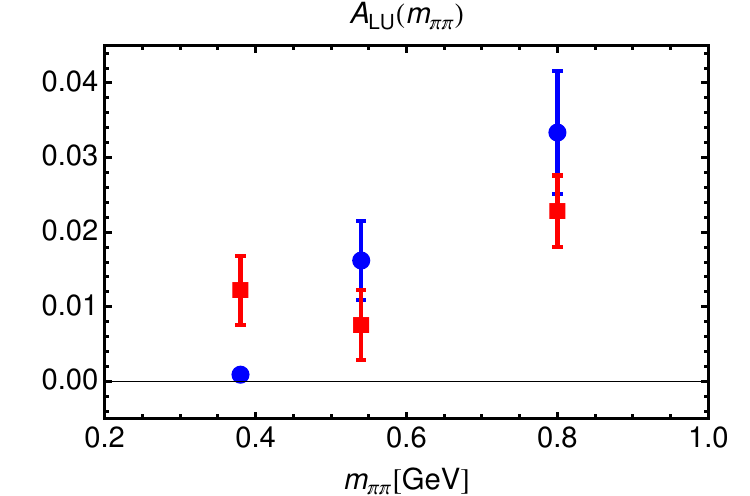}
\end{center}
\caption{1D projections of the BSA, repectively, in $z$ and $m_{\pi\pi}$. The red squares are the preliminary data of Ref.~\cite{BSA_silvia}. For the $z$ and the $m_{\pi\pi}$ projections, the blue circles represent the estimate for BSA Eq.~(\ref{eq:mhz}) from the DiFF fit~\cite{Courtoy:2012ry} for an integrated $x$-dependence, $n_x=0.21$.}
\label{fig:triptic_zmh}
\end{figure}

To check the presence of a possible twist-3 DiFF contribution, we have tried to reproduce  the $(z, m_{\pi\pi})$-dependences with the DiFF fits of Ref.~\cite{Courtoy:2012ry}. In the approximation of $D_1^u=D_1^d$ and neglecting the strange quark contributions,  we can write each projection as:
\begin{eqnarray}
A_{LU, \mbox{\tiny fit}}^{\sin \phi_R } \left(  x_i,  m_{\pi\pi, i},  z_i; Q_i, y_i \right)
&=&-\frac{W(y_i)}{A(y_i)}\,\frac{M}{Q_i}\,n_x \;  \, 
\frac{  \int_{z_{\text{\tiny min},\,i}}^{z_{\text{\tiny max},\,i}} dz \,\int_{(m_{\pi\pi, \, \text{\tiny min}})_{i}}^{(m_{\pi\pi, \, \text{\tiny max}})_{i}} \frac{|\bf{R}  |}{ m_{\pi\pi} }
 H_{1}^{\sphericalangle, u}\left( z,  m_{\pi\pi}, Q_i^2\right)  } 
       {   \int_{z_{\text{\tiny min},\,i}}^{z_{\text{\tiny max},\,i}} dz \,\int_{(m_{\pi\pi, \, \text{\tiny min}})_{i}}^{(m_{\pi\pi, \, \text{\tiny max}})_{i}}  D_{1}^u \left( z,  m_{\pi\pi}, Q_i^2\right) }\quad ,\nonumber\\
 \label{eq:mhz}
\end{eqnarray}
where the respective values of bins for the $z$ projections and the $m_{\pi\pi}$'s are given in Table.~\ref{tab:z_mh_1D}.
Within that approximation, the $x$-dependence is then only a scaling factor, 
\begin{eqnarray}
n_x=\frac{\int_{x_{\text{\tiny min}}}^{x_{\text{\tiny max}}} dx \,e^V(x, Q^2) }{\sum_q e^2_q\int_{x_{\text{\tiny min}}}^{x_{\text{\tiny max}}} dx f^{q+\bar{q}}(x, Q^2)}\quad,
\end{eqnarray}
 that in principle depends on $Q_i^2$ and on the interval $[x_{\text{\tiny min},\, i}, x_{\text{\tiny max},\, i}]$. This number is not known, but is related to the scale of the 1D projections. We show the result on Fig.~\ref{fig:triptic_zmh} for  $n_x=0.21$, value estimated in the rescaling of the fitting predictions Eq.~(\ref{eq:mhz}) {\it w.r.t.} the data. 
 
 The  predictions from the DiFFs fits are compatible, within error bars, with the preliminary data. The descrepancy observed in the low $m_{\pi\pi}$ behavior could be due to the limited range of validity of the DiFF fits. However, since the behavior of $\widetilde{G}^{\open}$ is not known, no real conclusion can be driven.
%

\begin{table}
 \centering
  \begin{tabular}{|>{\centering} m{1.cm} | |>{\centering} m{1.2cm} | >{\centering} m{2cm} | >{\centering} m{2.1cm} | >{\centering} m{2.1cm} |  >{\centering} m{1.8cm} | >{\centering} m{2.cm} | >{\centering} m{2.cm} |  m{1.8cm} | }
  \hline
bin $\#$ &	$x$   	&	$\langle Q^2\rangle$ [GeV$^2$]	& $z$ 	&	$m_{\pi\pi}$ [GeV] &	$n_u^{\uparrow}/n_u (Q^2)$   		 &    $ e^{V}_{\mbox{\tiny WW}}(x, Q^2)$ &    $ e^{V}_{\mbox{\tiny lead.}}(x, Q^2)$ 	&	$\Delta\left(e^{V}(x)\right)$ \\
 \hline
1&	0.171	&    	1.24		&	$[0.19,0.95]$	&	$[0.28, 1.66]$ &	0.199	&     	 2.611 	&     	 -0.263 	& 	0.578\\
2&	  0.244     &    	1.60 		&	$[0.20,0.95]$	&	$[0.28, 1.50]$ &	0.201    	&   	 0.687  	&     	 -0.850 	& 	0.238\\
3&	  0.356     & 	2.27 		&	$[0.21,0.92]$	&	$[0.28, 1.38]$ &	0.203   	 & 	 0.271	&     	 -0.243	 & 	0.091		\\
 \hline
 \end{tabular}
 \caption{The ratio of the integrated DiFFs and  the corresponding value for the flavor combination $e^{V}(x)=(4 e^{u_V}(x)-e^{d_V}(x))/9$. Note that the effect of evolution is of 1-2$\%$ at most when considering the ratio $n_u^{\uparrow}/n_u$ as the values of $Q_i^2$ are low. }
  \label{tab:res}
  \label{tab:resall}
\end{table}

\subsection{Beyond the WW scenario}

For completeness, we also consider the case in which  the  twist-3 DiFF $\widetilde{G}^{\open}$  is non-zero. We consider that it is of the order of magnitude of $\widetilde{D}^{\open}$.  As mentionned above, the preliminary results of CLAS~\cite{DSA_sergio} indicate that the $\cos\phi$ modulation of the DSA is small {\it w.r.t} leading-twist contributions. 

\begin{figure}[t]
\begin{center}
		\includegraphics[scale= .7]{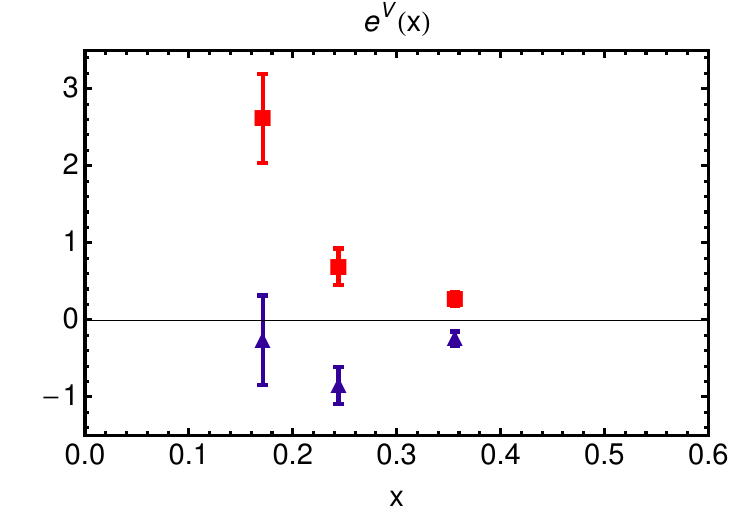}
\end{center}
\caption{Extractions on the combination $e^V\equiv4 e^{u_V}(x_i, Q_i^2)/9-e^{d_V}(x_i, Q_i^2)/9$. The red squares correspond to the {\it WW scenario}, the blue triangles to the {\it leading scenario}. The error bars correspond to the propagation of the experimental and DiFF errors. }
\label{fig:e_extract_us}
\end{figure}

The crucial observation is that the order of magnitude of $\widetilde{D}^{\open}$~\footnote{ $n_u^{\widetilde{D}^{\open}}(Q_i^2)$, integrated as  Eqs.~(\ref{e:nq},\ref{e:nqup}).} necessary to reproduce an integrated DSA is of a few percent of the integrated $H_1^{\open}$, Eq.~(\ref{e:nqup}). Within the present assumption, it translates in:
\begin{eqnarray}
n_u^{\widetilde{G}^{\open}}(Q_i^2) &\overset{\mbox{\tiny assump.}}\equiv&  n_u^{\widetilde{D}^{\open}}(Q_i^2) \cong \kappa \,n_u^{\uparrow}(Q_i^2)\quad,
\end{eqnarray}%
with $\kappa\sim0.2$ as estimated from the DSA of Ref.~\cite{DSA_sergio}. Then the BSA~(\ref{eq:alu}) becomes
\begin{eqnarray}
A_{LU, \mbox{\tiny leading}}^{\sin \phi_R } \left( x_i, m_{\pi\pi,\, i},  z_i ; Q_i, y_i\right) 
&=&-\frac{W(y_i)}{A(y_i)}\,\frac{M}{Q_i}\,\,\frac{\left[x_i e^V(x_i, Q_i^2) +\, \kappa f_1^V(x_i, Q_i^2)/z_i\right]\, n_{u,\, i}^{\uparrow}(Q_i^2)   } 
       { \sum_{q=u,d,s}\, e_q^2\,f_1^q(x_i , Q_i ^2)\, n_{q, \, i} (Q_i^2)  }\quad .\nonumber\\
       \label{eq:alu_lead}
\end{eqnarray}
Since the $(z, m_{\pi\pi})$-dependence is integrated,  a non-zero twist-3 PDF becomes manifest in  deviations from the trend in $x$ given by the unpolarized PDF contribution:
\begin{eqnarray}
A_{LU}^{\sin \phi_R }(x_i ; Q_i^2)\propto \frac{  \left(4\,f_1^{u_V}-  f_1^{d_V}\right)(x_i, Q_i^2)}{\left(4\, f_1^{u+\bar{u}}+\,f_1^{d+\bar{d}}\right)(x_i, Q_i^2)}\quad,
\label{eq:f1x}
\end{eqnarray}
the trend and size  of which can be estimated, {\it e.g.} with the MSTW08LO set.
Therefore, going beyond the WW approximation, the BSA  is straightfowardly inverted to get
\begin{eqnarray}
&&x_i^2\, e^{V}_{\mbox{\tiny lead.}}(x_i, Q_i^2)= -\frac{A(y_i)}{W(y_i)}\frac{Q_i}{M}\, A_{LU}^{\sin \phi_R} \left( x_i,  m_{\pi\pi,\,i},  z_i ; Q_i^2, y_i\right)\nonumber\\
&&\frac{1}{9} \,\frac{ 4 x_i f_1^{u+\bar u} (x_i, Q_i^2)\,n_{u, i}(Q_i^2)+  x_i f_1^{d+\bar d}(x_i, Q_i^2)\, n_{d, i}(Q_i^2)+  x_i f_1^{s}(x_i, Q_i^2)\, n_{s, i}(Q_i^2)}{n_{u, i}^{\uparrow}(Q_i^2)}\nonumber\\
&&-\kappa\,\frac{x_i}{ z_i}f_1^{V}(x_i, Q_i^2)
\quad,
\label{eq:therealextractBEYOND}
\end{eqnarray}
with  $\kappa=0.2$. In other words, the $x$-dependence coming from $f_1$ changes the extracted results of Fig.~\ref{fig:e_extract_us_ww}. 
The results are given in Tab.~\ref{tab:resall} for both scenarios and illustrated on Fig.~\ref{fig:e_extract_us}.Would the BSA only contain a contribution from $f_1 \widetilde{G}^{\open}$, the  twist-3 PDF $e(x)$, as extracted in this scenario, would be a constant.\footnote{A constant which value would be related to the uncertainty on $\kappa$.} 
 Our result shows that it is not the case. Hence, the behavior in $x$ of the BSA cannot be reproduced by the unpolarized PDF, Eq.~(\ref{eq:f1x}). We interpret this result as the first evidence for a non-zero $e(x)$ in the  range $x\in[0.126, 0.609]$.


A theoretical error on the extracted PDF could be  estimated to be the difference between the two scenarios, {\it i.e.}  varying $\kappa=0$ to $\kappa=0.2$. These two extreme scenarios set the constraints on the twist-3 PDF $e(x)$.
\\
%

\begin{table}[t]
 \centering
  \begin{tabular}{|>{\centering} m{1.3cm} | |>{\centering} m{2cm} | >{\centering} m{2cm}  | c| |>{\centering} m{2cm} | |>{\centering} m{2cm} | >{\centering} m{2cm} | c | }
  \hline
$z$-bin $\#$ &	$z$   &	$m_{\pi\pi}$ [GeV] 	&	$\langle Q^2\rangle$ [GeV$^2$] & $m_{\pi\pi}$-bin $\#$ &	$z$   &	$m_{\pi\pi}$ [GeV] 	&	$\langle Q^2\rangle$ [GeV$^2$]	\\
\hline
1&		$[0.19,0.45]$	&	$[0.28, 1.]$    &		1.74	&		1&		$[0.19,0.94]$	&	$[0.28, 0.46]$    &		1.81	\\
2&	  	$[0.46,0.65]$	&	$[0.28, 1.36]$ &	1.78	&		2&		$[0.23,0.94]$	&	$[0.48, 0.66]$    &		1.77    	\\
3&	 	$[0.66,0.95]$	&	$[0.28, 1.66]$ &	1.74	&		3&		$[0.31,0.95]$	&	$[0.66, 1.66]$    &		1.69   	 \\
 \hline
 \end{tabular}
 \caption{Binning, respectively, for the $z$ 1D projection and for the $m_{\pi\pi}$ 1D projection. }
  \label{tab:z_mh_1D}
\end{table}

\begin{figure}
\begin{center}
		\includegraphics[scale= .7]{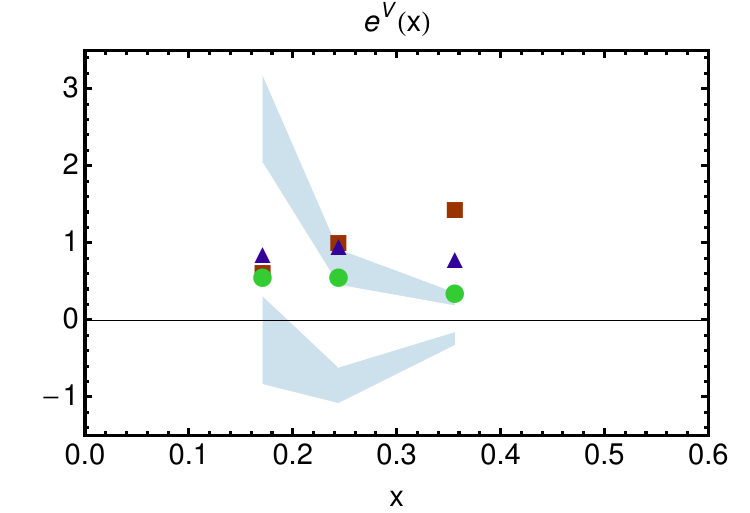}
\end{center}
\caption{Model predictions for combination $e^V\equiv4 e^{u_V}(x_i, Q_i^2)/9-e^{d_V}(x_i, Q_i^2)/9$ at the model scale $Q_0^2$, compared to the  error bands for the two scenarios  (light blue).  The blue triangles correspond to the bag model~\cite{Jaffe:1991ra,Avakian:2010br}, the red squares  to the spectator model~\cite{Jakob:1997wg} and the green circles the chiral quark soliton model~\cite{Ohnishi:2003mf}.}
\label{fig:e_extract_model}
\end{figure}

Thought the evolution of the twist-3 PDF is usually not applied in models, in Fig.~\ref{fig:e_extract_model} we propose an interpretive comparison with three --standard-- model predictions, {\it e.g.} the MIT bag model~\cite{Jaffe:1991ra}, the spectator model~\cite{Jakob:1997wg} and the chiral quark soliton model~\cite{Ohnishi:2003mf}.
The hadronic scale of  models symbolize the scale at which the model mimicks QCD for a given partonic representation, {\it e.g.} its partonic content. To 
purely valence quark models correspond a scale $Q_{0}^2\sim 0.1$ GeV$^2$ for $\Lambda_{\mbox{\tiny LO}}=0.27$ GeV~\cite{Gluck:1994uf,Traini:1997jz}. The author of~\cite{Wakamatsu:2000fd} refers to a scale $Q_0^2=0.3$ GeV$^2$.  
\\

\section{Conclusions}

We have presented the  extraction of chiral-odd PDF  $e(x)$ using the preliminary data for the Beam Spin Asymmetry in di-hadron SIDIS off proton target at CLAS.   The asymmetry consists in 2 terms: the first involves the twist-3 PDF of interest multiplied by a twist-2 DiFF and the second involves the usual unpolarized PDF multiplied by an unknown twist-3 DiFF. We have considered  two extreme scenarios:  a scheme where twist-3 DiFFs do not contribute and a second scenario in which the twist-3 DiFF is non-zero. While there are still non-negligible theoretical as well as experimental uncertainties, we show that the trend in Bjorken-$x$ cannot be reproduced by the PDF $f_1(x)$. It is an experimental evidence for a non-vanishing $e(x)$.

We have studied the BSA dependence in the DiFF variables and show that the trend in $z$ is compatible, within error bars, with the data. The behavior in $m_{\pi\pi}$ is not clear. Detailed studies of DiFF should be done in the future to enlarge the range of validity for the fits, improve the low energy functional form through a better $Q^2$ spanning of the data. Meanwhile, we cannot conclude on the presence of the twist-3 DiFF.

\vspace{1cm}

This manuscript has been prepared in collaboration with H.~Avakian, M.~Mirazita and S.~Pisano.
We are grateful to  A.~Bacchetta,  D.~Hasch, M.~Radici, P.~Schweitzer and M.~Wakamatsu for useful discussions.  This work was funded by the Belgian Fund F.R.S.-FNRS via the contract of ChargŽe de recherches (A.C.).
\\

%
\bibliographystyle{apsrevM}
\bibliography{alu_bib}
\end{document}